\documentclass[intlimits,twoside,a4paper]{article}
\usepackage[cp1251]{inputenc}
\usepackage[eqsecnum]{cmpj3}
\usepackage{bm} 

\issue{2023}{26}{1}{13506}
\doinumber{10.5488/CMP.26.13506}

\title[An Abelian Higgs model for disclinations in nematics]  {An Abelian Higgs model for disclinations in nematics}

\author[A. de P\'adua Santos, F. Moraes, F. A. N. Santos, S. Fumeron]{A. de P\'adua Santos\orcid{0000-0003-1262-0875}\refaddr{label1}, F. Moraes\orcid{0000-0001-7045-054X}\refaddr{label1}, 
	F. A. N. Santos\refaddr{label2,label3}, S. Fumeron\orcid{0000-0002-3429-0304}\refaddr{label4}\thanks{Corresponding author: \email{sebastien.fumeron@univ-lorraine.fr}.}}
\addresses{
\addr{label1} Departamento de F\'isica, Universidade Federal Rural de Pernambuco,
52171-900, Recife, PE, Brazil
\addr{label2} Dutch Institute for Emergent Phenomena (DIEP), Institute for Advanced Studies, University of Amsterdam, Oude Turfmarkt 147, 1012 GC, Amsterdam,  Netherlands
\addr{label3} Korteweg-de Vries Institute for Mathematics, University of Amsterdam, Science Park 105-107, 1098 XG Amsterdam,  Netherlands
\addr{label4} Laboratoire de Physique et Chimie Th\'eoriques, UMR Universit\'e de Lorraine - CNRS 7019, 54000  Nancy, France
}

\Keywords{analog gravity, liquid crystals, topological defects}

\date{Received November 20, 2022, in final form December 27, 2022}

\begin{document}

\maketitle

\begin{abstract}
Topological defects in elastic media may be described by a geometric field akin to three-dimensional gravity. From this point of view, disclinations are line defects of zero width corresponding to a singularity of the curvature in an otherwise flat background. On the other hand, in two dimensions, the Frank free energy of a nematic liquid crystal may be interpreted as an Abelian Higgs Lagrangian.  In this work, we construct an Abelian Higgs model coupled to ``gravity'' for the nematic phase, with the perspective of finding more realistic disclinations. That is, a cylindrically symmetric line defect of finite radius,  invariant under translations along its axis. Numerical analysis of the  equations of motion indeed  yield a  $+1$ winding number ``thick'' disclination. The defect is described jointly by the gauge and the Higgs fields, that compose the director field, and the background geometry. Away from the defect, the geometry is conical, associated to a dihedral deficit angle. The gauge field, confined to the defect, gives a structure to the disclination while the Higgs field, outside, represents  the nematic order. 
\printkeywords
\end{abstract}


\section{\label{intro}Introduction}
In a seminal paper \cite{kibble1976topology}, Tom Kibble proposed a mechanism to account for the formation of topological defects during phase transitions in the early Universe. Indeed, as a result of spontaneous symmetry breaking of gauge symmetry groups, cosmic line defects known as Nambu-Goto are expected to appear when the phase of the Higgs field presents singularities. These defects are infinitely thin and have no core structure. Surprisingly, the Kibble mechanism turned out to be also true for the formation of integer winding number disclinations in nematics and several works took advantage of this analogy to test Kibble mechanism from condensed matter experiments \cite{chuang1991cosmology, bowick1994cosmological}.

\begin{figure}[!htb]
    \centering
    \includegraphics[width=8cm]{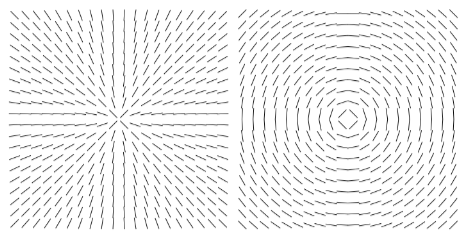}
    \caption{Representation of the average orientation of the molecules in the radial and circular winding number +1 disclinations.}
    \label{discl}
\end{figure}

Another well-known example of broken symmetry leading to the generation of defects is the Ginzburg-Landau theory of superconductivity \cite{de2018superconductivity}, where the superconductivity order parameter  couples to the magnetic field. The broken symmetry superconducting transition includes Abrikosov's vortices that encapsulate the higher symmetry (normal) phase within the broken symmetry (superconducting) phase. Similarly, the isotropic-nematic transition in liquid crystals, ruled by the Frank free energy, leads to the creation of disclinations. Like Abrikosov's vortices, $+1$ winding number disclinations encapsulate the higher symmetry (isotropic) phase within the broken symmetry (nematic) phase.    Perhaps  not so well-know is the fact that the Frank free energy can be expressed as an Abelian Higgs Lagrangian \cite{kurz2000hydrodynamics}. That is, an Abelian gauge field coupled to the Higgs (the notorious Mexican hat) potential.
In this work, we bring together this Abelian Higgs description of the nematic phase and the geometric theory of defects (described briefly in the next section) like it was done for cosmic strings  by Garfinkle \cite{garfinkle1985general} and many others that followed. Our interest here is to apply the above ideas to the study of the winding number +1 disclinations depicted in figure \ref{discl}.  As can be easily inferred from the figure, there should be a great cost in elastic energy to have singularities at the defect center. Indeed, the elastic energy involved is so high that it is reasonable to assume that the defect core consists of a region of a more disordered higher energy phase. Depending on the model and type of disclination, the core phase may be isotropic, escape into the third dimension or biaxial.  In any case, one should expect a smooth transition in the order parameter from the center of the core to the outer regions.  This is evident in our approach, where the director, as represented by the Higgs field, presents the behavior shown in figure \ref{fig2}. At the same time, the disclination core is threaded by an Abelian flux which represents the higher energy phase order trapped inside the defect. 

\section{\label{model}The model}
\subsection{Topological defects in elastic media and 3D gravity}
{Cosserat media form a particular class of elastic materials as they display a spin structure. They consist of a continuous collection of particles that can not only move, but also display rotational degrees of freedom \cite{cosserat1896theorie}: they can, therefore, transmit both forces and torques to each other. This is, for instance, the case of spin glasses and nematic elastomers, which are  elastic solids with a spin structure. A disclination is a topological defect that corresponds to a line singularity of the director field orientation.}

{In \cite{katanaev1992theory}, Katanaev and Volovich used Riemann-Cartan manifolds to describe static disclinations: the surface density of Frank vector field is translated as a curvature of the background geometry on which low-energy elastic excitations propagate. Besides its mathematical elegance (in ordinary elasticity, defects are introduced from a complicated set of boundary conditions), this geometric theory of defects has two assets: 1) it offers  an intuitive insight on the impact of a defect onto transport phenomena and 2) it provides with a Taylor-made mapping between defects in condensed matter and defects in cosmology (cosmic strings).}

In the Katanaev and Volovich model, topological defects in elastic media are solutions to the Lagrangian
\begin{equation}
    L = \kappa R - 2\gamma R_{[ij]}R^{[ij]},
\end{equation}
where $\kappa$ and $\gamma$ are dimensioning parameters, and the elastic deformation is represented by the Ricci curvature $R$. The antisymmetric part of the Ricci tensor $R_{[ij]}=\frac{1}{2}(R_{ij}-R_{ji})$ introduces torsion as required for the existence of dislocations. In the absence of these, we are left with the Hilbert-Einstein Lagrangian $L=\kappa R$ in 3D.

\subsection{Frank free energy as an Abelian Higgs Lagrangian}
\begin{figure}[!htb]
\begin{center}
\includegraphics[width=3cm,height=4cm]{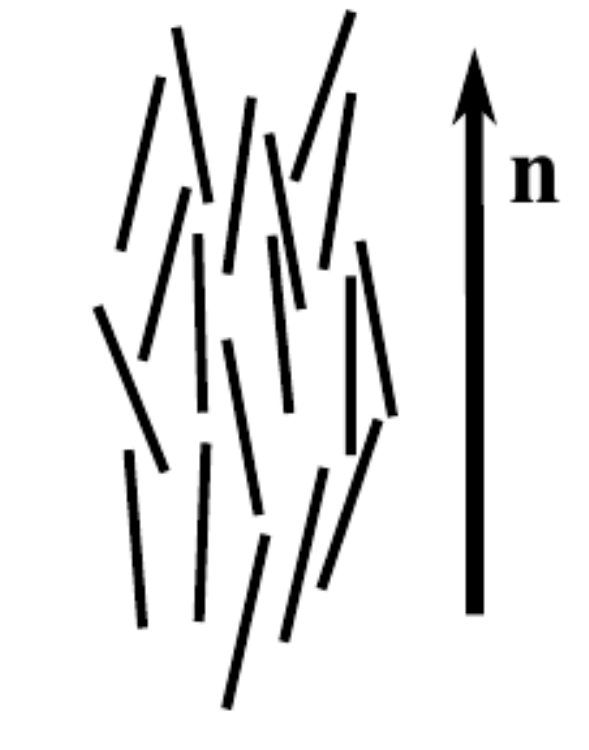}
\caption{The average local orientation of nematic molecules is specified by the director field $\bf n$.}
\label{fig1}
\end{center}
\end{figure}

The nematic phase of liquid crystals is characterized by the director field $\bf n$, which defines the local average orientation of the molecules as shown in figure \ref{fig1}.
The static configuration of the nematic liquid crystals can be described by the Frank free energy density \cite{de1993physics},
\begin{equation}
\mathcal{F}  = \frac{1}{2} \bigg[K_1(\nabla\cdot \vec{n})^2+K_2(\vec{n}\cdot\nabla\times \vec{n})^2
   +K_3(\vec{n}\times\nabla\times \vec{n})^2\bigg], \label{free1}
\end{equation}
where  $K_1, K_2, K_3$ are elastic constants associated respectively with splay, twist and bend distortions in the liquid crystal. If one considers the approximation of one elastic constant $K_1 = K_2 = K_3 = K$, the free energy (\ref{free1}) reads \cite{de1993physics}
\begin{equation}
\mathcal{F} = \frac{1}{2}K\partial_l n_i\partial_l n_i .
\label{free2}
\end{equation}
Since the winding number $+1$ disclinations have planar $\vec{n}$, we define the complex director field 
\begin{equation}
    n=n_1 + \ri n_2 .
\end{equation}
With this, equation \eqref{free2} can be written in a gauge covariant form as \cite{kawasaki1985gauge}
\begin{equation}
  \mathcal{F} = \frac{1}{2}K D_j n \overline{D_j n}  + \frac{1}{2}F^2 , \label{free4}
\end{equation}
where the covariant derivative $D_j=\partial_j -i  A_j$ and $F=\partial_1 A_2 - \partial_2 A_1$. The bar denotes complex conjugation and the term in $F^2$ is included to make the gauge field $A_i$ a dynamical variable. The gauge field dependence on the complex director $n$ is given by  \citep{yi} 
\begin{equation}
A_i = \varepsilon^{jk}n^{j}\partial_i n^{k}, 
\end{equation}
where $\varepsilon^{jk}$ is the two-dimensional Levi-Civita symbol.

Topological defects can be considered as small regions where the order is not well defined. Thus, in order to model such regions we relax the usual boundary condition $|{n}|=1$ allowing the size of the director to vary between 0 (in the center of the defect) and 1 (away from the defect). To do this we use a Higgs-like potential 
\begin{equation}
V(n) = \frac{\lambda}{4}(|n|^2 - 1)^2 , \label{pot}
\end{equation}
which is minimized at $|{n}|=1$
This way, the interior of the defect is modelled by an isotropic fluid ($|{n}|=0$) and the exterior is modelled by the nematic phase ($|{n}|=1$) corresponding to the ground state of \eqref{pot}. The ratio $K /\lambda$  is proportional to the core size of the disclination.  By adding the potential \eqref{pot} to the free energy density~\eqref{free4} we obtain \cite{kurz2000hydrodynamics}
\begin{equation}
  \mathcal{L}_m = \frac{1}{2}K D_j n \overline{D_j n}  + \frac{1}{2}F^2 +\frac{\lambda}{4}(|n|^2 - 1)^2 . \label{free5}
\end{equation}

\subsection{Coupling elasticity to the orientation field}
Matter fields in curved spacetime are generally described by the action
\begin{equation}
 S = \int \rd^4x \sqrt{-g}\left(\frac{1}{16\piup G}R + \mathcal{L}_m\right), \label{Action}
\end{equation}
 where $R$ is the Ricci scalar and $G$ denotes Newton's constant. $ \mathcal{L}_m $ is the matter Lagrangian density. Both terms are usually coupled via the so-called minimum coupling, where  the geometry enters the covariant derivatives that replace the ordinary ones from flat space.

In the case studied here, we have
\begin{equation}
 S = \int \rd^3x \sqrt{-g}\left(\kappa R + \mathcal{L}_m\right), \label{eqAction}
\end{equation}
where  $\mathcal{L}_m$ is given by equation \eqref{free5}.

\subsection{The ansatz}
\label{Ansatz}
Let us consider a line defect placed on the $z$ axis, long enough as to allow us to assume translational invariance along $z$. In order to benefit from the numeric machinery available in similar calculations, we choose to work in four-dimensional spacetime, keeping in mind that the temporal dimension cannot be affected by the relevant fields. This said, we consider the most general, 
cylindrically symmetric line element invariant under boosts  along the $z$-direction. In three-dimensional space, this reduces to the required translational invariance.
By using cylindrical coordinates, this line element is given by:
\begin{equation}
 \rd s^2 = c^2 N^2 (\rho) \rd t^2 - \rd\rho^2 - L^2(\rho)\rd\phi^2 - N^2(\rho)\rd z^2 \  ,
 \label{ds}
\end{equation}
where $N(\rho)$ and $L(\rho)$ are functions of $\rho$ only.
As we shall see below, the numerical results indicate that $N^2(\rho)$ is essentially constant and equal to one, ensuring that the time dimension is not affected.
For this metric, the only non-vanishing components of the Ricci tensor, $R_{\mu\nu}$,  are:
\begin{equation}
 R_{tt} = - R_{zz} = \frac{NLN''+ NN'L' + L(N')^2}{L} \  ,
\end{equation}
\begin{equation}
 R_{\rho\rho} = \frac{2LN'' + NL''}{NL} \  ,
\end{equation}
\begin{equation}
 R_{\phi \phi} = \frac{L(2N'L' + NL'')}{N}   ,
\end{equation}
where the primes denote derivatives with respect to $\rho$.

For the Higgs and gauge fields, we use the expressions below:
\begin{equation}
n =  X(\rho)\re^{\ri m\phi},
\end{equation}
\begin{equation}
\vec{A}(\rho) = \hat{\phi}\left[{m-H(\rho)}\right] =
 -\hat{\phi}{A(\rho)},  
\end{equation}
where $m$ is the winding number. Particularly, in this work we are interested in $m=1$.

\section{Equations of motion and boundary conditions}
\label{Equation}

For convenience of the  numerical analysis, we make $K=1$ in equation \eqref{free5} and
define 
\begin{eqnarray}
 x & = &\sqrt{\lambda}\rho,\nonumber \\
 L(x) & = & \sqrt{\lambda} L(\rho).
\end{eqnarray}
Therefore, the Lagrangian now depends only on two  parameters: $\kappa$ and $\lambda$. 

Varying the action (\ref{eqAction}) with respect to the matter and the metric fields, we obtain a system of
four non-linear coupled differential equations. These are the Euler-Lagrange equations:
\begin{equation}
 \frac{(N^2LX')'}{N^2L}= X\left[X^2-1 + \frac{H^2}{L^2}\right] \  , \label{eq1}
\end{equation}

\begin{equation}
 \frac{L}{N^2} \left(\frac{N^2H'}{L}\right)' = \frac{1}{\lambda} X^2H  \  .  \label{eq3}
\end{equation}

\begin{equation}
 \frac{\left(LNN'\right)'}{N^2L} = \frac{1}{2\kappa} \left[\frac{\lambda H'^2}{2 L^2} -\frac{1}{4}\left(X^2-1\right)^2 \right],  \label{eq4}
\end{equation}
and
\begin{equation}
\frac{\left(N^2L'\right)'}{N^2L} = -\frac{1}{2\kappa} \left[\frac{\lambda H'^2}{2 L^2} + \frac{X^2H^2}{L^2}+\frac{1}{4}\left(X^2-1\right)^2  
 \right] .  \label{eq5}
\end{equation}
The primes in the above equations denote derivatives with respect to $x$.

The requirements of regularity at the origin and topologically stable solutions,  lead to the following boundary conditions 
for the matter and gauge fields
\begin{equation}
     H(0) = 1; \quad H(\infty) = 0 \  , \label{eqbound1}
\end{equation}
\begin{equation}
     X(0) = 0, \quad X(\infty) = 1,   \label{eqbound2}
\end{equation}
and for the metric fields
\begin{equation}
     N(0) = 1, \quad N'(0) = 0 \  , \quad L(0) = 0, \quad L'(0) = 1 \  . \label{eqbound3}
\end{equation}

\section{\label{mumres} Numerical results}

 In this section we  numerically analyze the behavior of the solutions of the system of
differential equations that describe the Abelian vortex for the choice of physical parameters of the system. To perform the numerical study, we used the computer program  COLSYS (Collocation Software for Boundary-Value ODEs) \cite{colsys1, colsys2, colsys3}. The system of differential equations is described
through a set of functions $u_i(x)$, where $i =1, 2, ..., d$ and $d \in [a, b]$ are written as
\begin{equation}
    \frac{\rd^{m_i}u_i}{\rd x^{m_i}} = f_i(x,z(u)),
\end{equation}
where $m_i$ is the order of the differential equation of the function $u_i$  and $z(u)$ is the set of functions and their derivatives given by
\begin{equation}
z(u) = \left[u_1, \frac{\rd u_1}{\rd x}, \dots, \frac{\rd u_1^{m_1-1}}{\rd x^{m_1-1}}, \dots, u_d, \frac{\rd u_d}{\rd x}, \dots, \frac{\d u_d^{m_d-1}}{\d x^{m_d-1}}\right].
\end{equation}

{This way, we  have a system of $d$ differential equations with $d$ functions to be determined. }

{The boundary conditions are written as
\begin{equation}
g_j(z(u(\eta_j)))=0,
\end{equation}
where $j= 1,\cdots, m^{\ast}$, such that $m^{\ast} = \sum_{i=1}^{d}m_i$ and $\eta_j \in [a, b]$, $\eta_1\leqslant\eta_2\leqslant\cdots\leqslant\eta_{m^{\ast}}$. These are the points
where differential equations must meet the boundary conditions. We  assume that
 the system has numerical solutions such that the functions $u_i(x) \in C^{m_i-1}$ are in the integration range.  COLSYS is considered to be state-of-the-art software for solving contour value problems that implement a series of algorithms to solve differential equations.}
 
\begin{figure}[!t]
\begin{center}
\includegraphics[width=8cm,height=7cm]{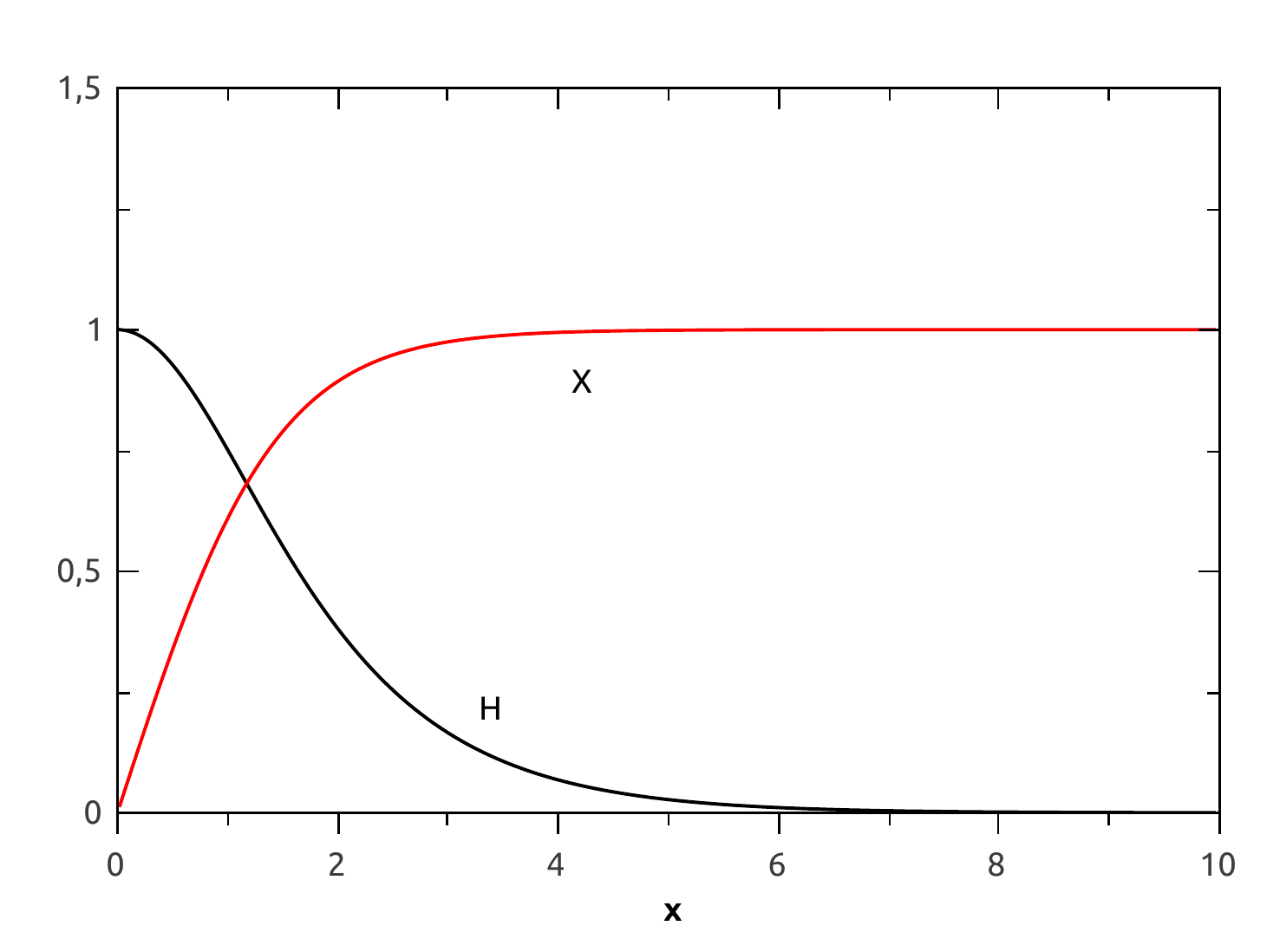}
\caption{(Colour online) The behavior for the Higgs $(X)$ and gauge $(H)$ fields as functions of $x$,
considering the parameters  $\lambda=1.0$, $\kappa=0.83$.}
\label{fig2}
\end{center}
\end{figure}

\begin{figure}[!b]
\begin{center}
\includegraphics[width=9.5cm]{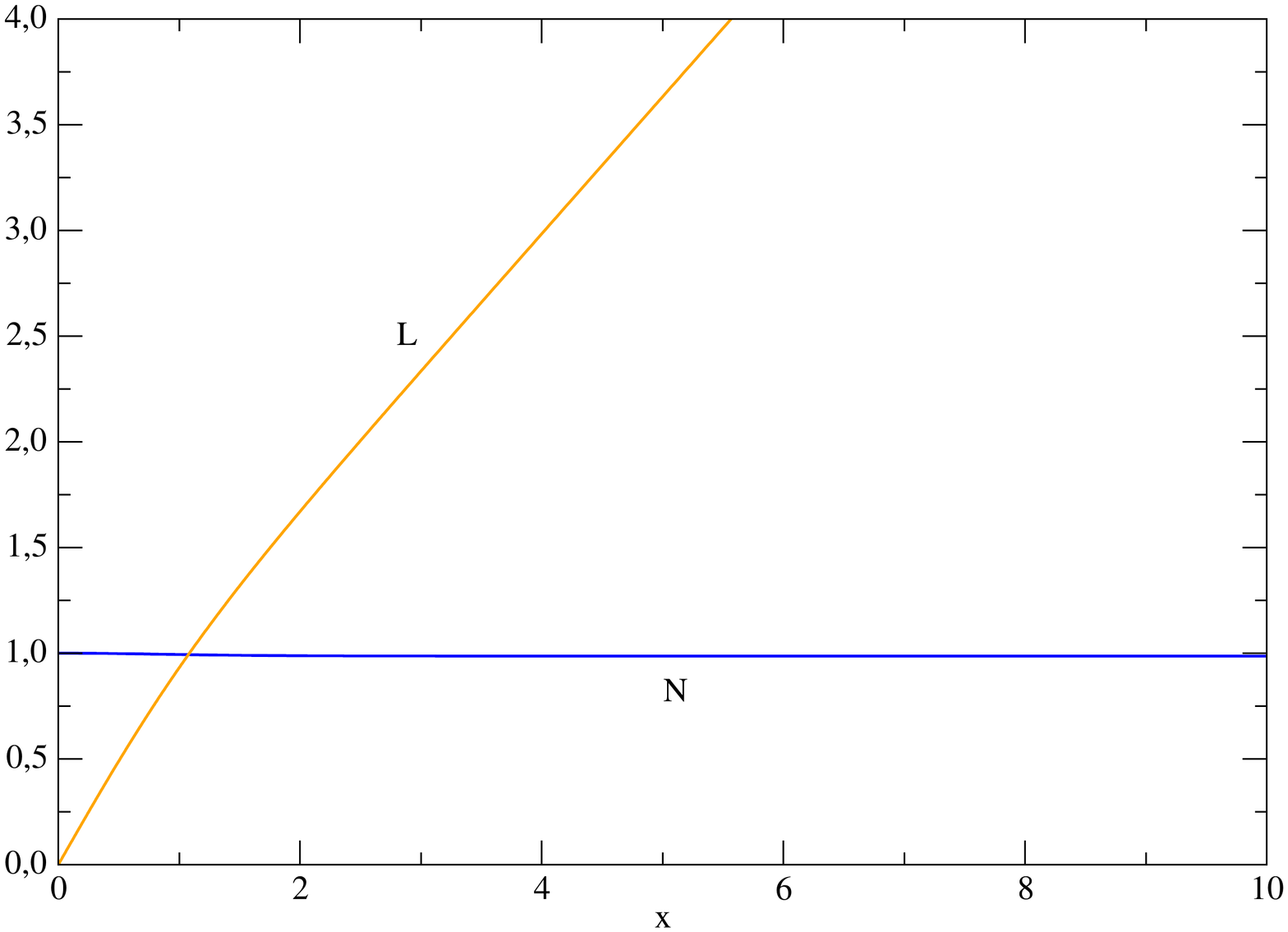}
\caption{(Colour online) The behaviour of the geometric fields $L$ and $N$ as functions
of $x$, considering the parameters  $\lambda=1.0$, $\kappa=0.83$.}
\label{fig3}
\end{center}
\end{figure}

 Our results are summarized by figures \ref{fig2} and \ref{fig3}. In the former, there is shown  the behaviour of the gauge ($H$) and Higgs ($X$) fields while in the latter we have the geometric fields $L$ and $N$.

\section{Discussion}

In figure \ref{fig2} we see the radial profiles of the gauge and the Higgs fields associated to the director. Their behavior is identical to the ones found by Garfinkle \cite{garfinkle1} for  Abelian cosmic strings. We also see the same behavior as the magnetic field and superconducting order parameter  predicted by the Ginzburg-Landau theory of type II superconductivity \cite{de2018superconductivity}. No surprises here, since the Lagrangians of all these models share the same form, except for the GL case where there is no geometric part.

Figure \ref{fig3} shows the geometric fields $L(x)$ and $N(x)$. Note the inflexion of $L(x)$ at the defect radius ($x\approx 1$ as seen in figure \ref{fig2}) signaling a change in the geometry as one moves out of the defect. Note also that we found  $N(x)=1$ justifying the 4D approach we used and, most importantly, that the time dimension is not affected.

The geometry associated to the $m=1$ disclination, as pointed out by Katanaev and Volovich, is conical. That is, in their purely geometric model, it corresponds to a space with a missing dihedral angle. This missing angle, or  deficit angle, means that a circle around the defect comprises a total angle less than $2\piup$. By analysing the slope of $L(x)$ away from the disclination core, we have for the deficit angle:
\begin{equation}
 \label{deficit}
 \delta=2\piup[1-L'(\infty)]  \  . 
\end{equation}
Using the parameter values of the  previous plots,
we find for the planar angle deficit 
\begin{equation}
\frac{\delta_A}{2\piup}\approx 0.3493 \  .
\end{equation}

\section{\label{concl}Perspectives}

 In 1987, Schopohl and Sluckin  \cite{schopohl1987defect} predicted that the core of half-integer winding number disclinations in a nematic liquid crystal is in a biaxial phase instead of being isotropic. This has been confirmed by computer simulations, as can be seen in  \cite{bhattacharjee2008numerical,de2010nematic,avelino2011electric,schimming2020anisotropic}, for instance. On the other hand, the Kibble mechanism has  been observed in the creation of  such defects  \cite{mukai2007defect}. Since biaxiality appears naturally when one deals with $\text{SO(3)}$, the non-Abelian symmetry group of the liquid crystal in 3D, this motivates us to extend the present model to the non-Abelian case, in the spirit of \cite{de2015gravitating} where a SU(2) Higgs model coupled with gravity was investigated to account for the formation of non-Abelian cosmic strings. This work is presently in progress. Further extensions of this work include the introduction of dynamics and the study of the Kibble-Zurek mechanism both in the Abelian and non-Abelian cases.


\ukrainianpart

\title[Абелева модель Хіггса для дисклінацій у нематиках]  {Абелева модель Хіггса для дисклінацій у нематиках}
\author{А. де Падуа Сантуш\refaddr{label1}, Ф. Мораеш\refaddr{label1}, 
	Ф. А. Н. Сантуш\refaddr{label2,label3}, С. Фумерон\refaddr{label4}}
\addresses{
	\addr{label1} Фізичний факультет, Федеральний сільський університет Пернамбуку,
	52171-900, Ресіфе, Бразилія
	\addr{label2} Голландський інститут надзвичайних явищ, Інститут перспективних досліджень, Університет Амстердама, Уде Турфмаркт 147, 1012 Амстердам, Голландія
	\addr{label3} Математичний інститут ім. Кортевега-де Фріза, Університет Амстердама, Науковий парк 105-107, 1098 Амстердам, Голландія
	\addr{label4} Лабораторія теоретичної фізики та хімії, Університет Лотарингії, CNRS, F-54000 Нансі, Франція
}

\makeukrtitle

\begin{abstract}
	Топологічні дефекти у пружному середовищі можна описати геометричним полем, спорідненим з тривимірною гравітацією.
	З цієї точки зору, дисклінації — це лінійні дефекти нульової ширини, що відповідають сингулярності кривизни загалом плоского фону. З іншого боку, у двовимірному випадку вільну енергію Франка нематичного рідкого кристала можна інтерпретувати як абелів лагранжіан Хіггса. У цій роботі ми будуємо абелеву модель Хіггса у поєднанні з ``гравітацією'' для нематичної фази та перспективою пошуку більш реалістичних дисклінацій.
	Таким чином, розглядається циліндрично-симетричний лінійний дефект скінченного радіуса, інваріантний щодо трансляцій уздовж своєї осі. Як показав числовий аналіз рівнянь руху, у системі справді виникає ``товста'' дисклінація з $+1$ номером намотування.  Дефект описується ка\-лібру\-валь\-ним полем та полями Хіггса, які формують поле директора, і фоновою геометрією.  Здаля від де\-фек\-ту геометрія є конічною, і вона пов’язана з двогранним кутом дисклінацiї. Калібрувальне поле, обмежене дефектом, задає структуру дисклінації, тоді як поле Хіггса ззовні представляє нематичний порядок.
	
	\keywords аналогова гравітація, рідкі кристали, топологічні дефекти
\end{abstract}

\lastpage
\end{document}